\documentclass[twocolumn,
notitlepage,
aps,
prx,
floatfix,
superscriptaddress,
nofootinbib
]{revtex4-2}

\usepackage{amsmath, amsfonts, amsthm, amssymb, bbm}
\usepackage[margin=1in]{geometry}
\usepackage{graphicx} 
\usepackage{xcolor}
\usepackage[colorlinks=true, hyperindex, breaklinks, linkcolor=blue, urlcolor=blue, citecolor=blue]{hyperref}

\usepackage{physics}
\usepackage{dcolumn}
\usepackage{microtype}
\usepackage{tabularx} 
\usepackage{enumitem} 
\usepackage{mathtools}
\usepackage{float}
\usepackage{bm}
\usepackage{subfiles}
\usepackage{tikz} 
\usetikzlibrary{tikzmark, arrows.meta}
\usepackage{algpseudocode}
\floatstyle{ruled}
\newfloat{algorithm}{t}{loa}
\floatname{algorithm}{Algorithm}
\makeatletter
\def\ftype@algorithm{4}
\newbox\float@algorithm@box
\expandafter\let\csname fbox@4\endcsname\float@algorithm@box
\setbox\float@algorithm@box\hbox{}
\makeatother

\algnewcommand\algorithmicforeach{\textbf{for each}}
\algdef{S}[FOR]{ForEach}[1]{\algorithmicforeach\ #1\ \algorithmicdo}
\newcommand{\PyComment}[1]{\ttfamily\textcolor{gray}{\# #1}}

\newcommand{\dcirc}{d_{\mathsf{circ}}}

\date{\today}

\begin{document}

\preprint{APS/123-QED}

\title{Stairway Codes: Floquetifying Bivariate Bicycle Codes and Beyond}

\author{Shoham Jacoby}
\affiliation{AWS Center for Quantum Computing, Pasadena, CA 91125, USA}
\affiliation{Racah Institute of Physics, The Hebrew University of Jerusalem, Jerusalem 91904, Givat Ram, Israel}
\author{Alex Retzker}
\affiliation{AWS Center for Quantum Computing, Pasadena, CA 91125, USA}
\affiliation{Racah Institute of Physics, The Hebrew University of Jerusalem, Jerusalem 91904, Givat Ram, Israel}
\author{Fernando Pastawski}
\affiliation{AWS Center for Quantum Computing, Pasadena, CA 91125, USA}

\begin{abstract}
  Floquet codes define fault-tolerant protocols through periodic measurement sequences that drive a dynamically evolving stabilizer group. 
  They provide a natural framework for hardware supporting two-qubit parity measurements but no unitary entangling gates.
  However, few known constructions achieve both high encoding rates and high thresholds. 
  We close this gap by introducing Stairway codes, a family of high-rate Floquet protocols obtained by Floquetifying Abelian two-block group algebra codes,
  a class that includes the bivariate bicycle codes. 
  By representing the static code as a foliated ZX-calculus network within a $(w{-}1)$-dimensional space-time lattice and rotating the time axis, we decompose its weight-$w$ stabilizers into a periodic sequence of pairwise measurements.
  This reduces the design of new codes within this family to the selection of favorable periodic boundary conditions. 
  We identify instances with competitive parameters, analyze their distance under circuit-level noise, 
  and demonstrate logical error rates surpassing those of other Floquet codes at comparable encoding rates. 
  Remarkably, our construction requires fewer than $300$ physical qubits to match the distance and encoding rate of semi-hyperbolic Floquet codes that use over $1300$ qubits.
\end{abstract}

\maketitle

\section{Introduction}

\begin{figure*}
\centering
    \def\svgwidth{\linewidth}
    {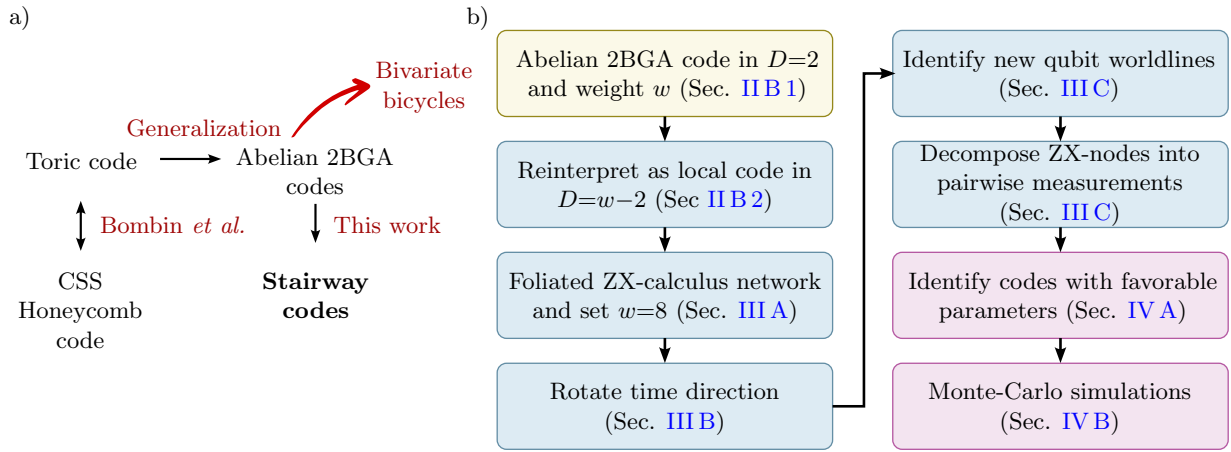}
    \caption{
    a) Outline of our results. We create a new construction from Abelian two-block group algebra (2BGA) codes using the same technique used in Ref.~\cite{Bombin_2024} to connect the toric code and the CSS honeycomb code.
    Abelian 2BGA codes~\cite{lin2023twoblockgroupalgebra} are a generalization of the toric code that includes the bicycle~\cite{kovalev2013bicycle_codes} and bivariate bicycle~\cite{Bravyi_2024BB_codes} code families.
    b) Outline of this work, with corresponding sections for each stage.
    }
    \label{fig:introduction}
\end{figure*}

Protecting quantum information from noise is a central challenge on the path to scalable quantum computing.
Quantum fault tolerance addresses this by encoding logical qubits into many physical qubits and repeatedly measuring parity checks to detect and correct errors~\cite{Preskill_1998, Terhal_2015}.
The practical viability of such schemes depends critically on the interplay between the number of physical qubits per logical qubit (encoding overhead), the number of errors the code can tolerate (code distance), and the complexity of the operations required by the hardware.

Floquet codes~\cite{Hastings_2021} are a paradigm for quantum fault-tolerance based on general periodic driving, which is not limited to the standard approach of repeated stabilizer measurements.
These admit a dynamically evolving stabilizer group and can naturally model circuits composed entirely of pairwise measurements.
This makes them particularly well-suited for architectures supporting these native operations, such as Majorana qubits~\cite{aguado2020majorana,aasen2025majorana}, distributed computing~\cite{de_Bone_2024distributed_surface_code}, and metamaterial-based long-range couplers for dual-rail qubits~\cite{golan2025}.

However, prominent Floquet implementations such as the honeycomb code~\cite{Hastings_2021, Gidney_2021Honeycomb} and the Calderbank--Shor--Steane (CSS) honeycomb code~\cite{Bombin_2024, Kesselring_2024anyons} encode only one or two logical qubits independent of block size.
Recent work on hyperbolic Floquet codes~\cite{Higgott_2024hyperbolic, fahimniya2025hyperbolicfloquet} has improved encoding rates, but these constructions have not matched the rates and distances achievable by the best static code families at comparable block sizes.
In particular, Abelian two-block group algebra (2BGA) codes~\cite{lin2023twoblockgroupalgebra,wang2023other_paper_2bga}, such as bivariate bicycle codes~\cite{Bravyi_2024BB_codes}, have set a higher standard for encoding rates and distances in finite-size blocks.

In this work, we close this gap by introducing Stairway codes: a new family of high-rate Floquet protocols derived from Abelian 2BGA codes.
Our construction generalizes the technique recently applied by Bombin et al.~\cite{Bombin_2024} to connect the toric code with the CSS honeycomb code (see Fig.~\ref{fig:introduction}a).
While the method applies to Abelian 2BGA codes in general, we focus here on bivariate bicycle codes, where the underlying group is a product of two cyclic groups.

Specifically, we treat a weight-$w$ Abelian 2BGA code as a local system in a $(w{-}2)$-dimensional space~\cite{arnault2025_2bga_boundaries} and represent its syndrome extraction circuit as a ZX-calculus network embedded in $(w{-}1)$ dimensions (See Fig.~\ref{fig:introduction}b). 
Within this network, we identify new qubit worldlines---the paths of physical qubits through the circuit---that are tilted relative to the original time axis.
By using ZX-calculus identities to rewrite the network, we effectively ``rotate'' the time direction within the lattice geometry, converting the static structure of Abelian 2BGA codes into a dynamic, pairwise measurement schedule.

Through a systematic numerical search over this family, we identify several instances with both high encoding rates and large code distances; we showcase some of these in Table~\ref{table:codes_intro}.
Stairway codes operate at significantly smaller block sizes than other high-rate Floquet codes with similar parameters, making them more amenable to near-term implementation.

We evaluate Stairway codes using the BPT metric $kd^2/n$~\cite{Bravyi_2010BPT_bound}. 
The standard honeycomb code is limited to a constant value of $1/3$ in this metric, 
while hyperbolic Floquet codes~\cite{vuillot2021planarfloquetcodes, Higgott_2024hyperbolic,fahimniya2025hyperbolicfloquet} achieve a scaling of $O(\log^2 n)$.
In this work, we focus on finite-size instances, where our numerical results demonstrate highly competitive values of this metric,
with substantial improvement over both families.

\begin{table*}[t]
\begin{center}
\begin{tabular}{c|c|c|c|c|c}
\hline
& & & & & \\
$[[n,k,d]]$ & \parbox{2.7cm}{\centering Net encoding \\ overhead $n/k$} & \parbox{2cm}{\centering BPT constant 
\\ $kd^2/n$} & \parbox{3cm}{\centering Circuit-level\\ distance $\dcirc$ \\ (EM3 noise model)} &
\parbox{2.7cm}{Pseudo-threshold \\ $p_0$ } & 
\parbox{2cm}{$p_L(0.001)$}\\
& & & & & \\
\hline
\hline
$[[192,16,4]]$ & $12$ & $1.33$ & $4$ & - & $4 \times 10^{-3}$ \\
\hline 
$[[192,8,6]]$ & $24$ & $1.5$ & $6$ & $2.5 \times 10^{-3}$ & $1 \times 10^{-5}$ \\
\hline
$[[288,14,9\le d\le10]]$ & $20.6$ & $\in\{3.94, 4.86\}$ & $\le 7$ & $2.2 \times 10^{-3}$ & $1 \times 10^{-5}$ \\
\hline
$[[384,16,\le12]]$ & $24$ & $\le 6$ & $\le 8$ & $2.1 \times 10^{-3}$ & $8 \times 10^{-6}$ \\
\hline
$[[576,26,\le12]]$ & $22.2$ & $\le 6.5$ & $\le8$ & - & - \\
\hline
$[[576,14,\le 20]]$ & $41.1$ & $\le 9.72$ & $\le14$ & $2.5 \times 10^{-3}$ & $\sim 3 \times 10^{-7}$ \\
\end{tabular}
\end{center}
\caption{Example of codes in the Stairway codes family. 
The distance $d$ given here is the \emph{embedded distance} (Sec.~\ref{sec:distance-results})~\cite{Hastings_2021,Higgott_2024hyperbolic}.
The encoding rate $r=k/n$ is also the net encoding rate, as ancillas are not required for these codes. 
The circuit level distances are computed using integer linear programming~\cite{landahl2011ilp_distance_color_code} decoder. When the program did not find an optimal solution, an upper bound is reported.
The pseudo-threshold and the logical error rate $p_L$ are presented for memory experiment of $d$ rounds using Monte-Carlo sampling.
The pseudo threshold is the break-even point, defined as the point at which $p_L=kp$. The symbol $\sim$ is used whenever a direct sampling was not achieved and the value is extrapolated from higher error rates.
The EM3 noise model~\cite{Gidney_2021Honeycomb} is used for both the circuit-level distance $\dcirc$ and logical error rate calculations $p_0$ and $p_L$.
}
\label{table:codes_intro}
\end{table*}

Using Monte Carlo simulations with the EM3 circuit noise model~\cite{Gidney_2021Honeycomb}, we benchmark Stairway codes against hyperbolic Floquet codes and show they achieve $10\times$ lower logical error rates at $p=10^{-3}$ for similar block sizes and better encoding rate.
We also compare against two compilations of bivariate bicycle syndrome extraction circuits into pairwise measurements, inspired by Refs.~\cite{Gidney_2023,rodatz2024floquetifying_distance_preserving}, and show that Stairway codes surpass both by a significant margin.

\section{Preliminaries}

\subsection{ZX-calculus and diagrammatic notation}

\begin{figure*}[t]
    \centering
    \def\svgwidth{\linewidth}
    {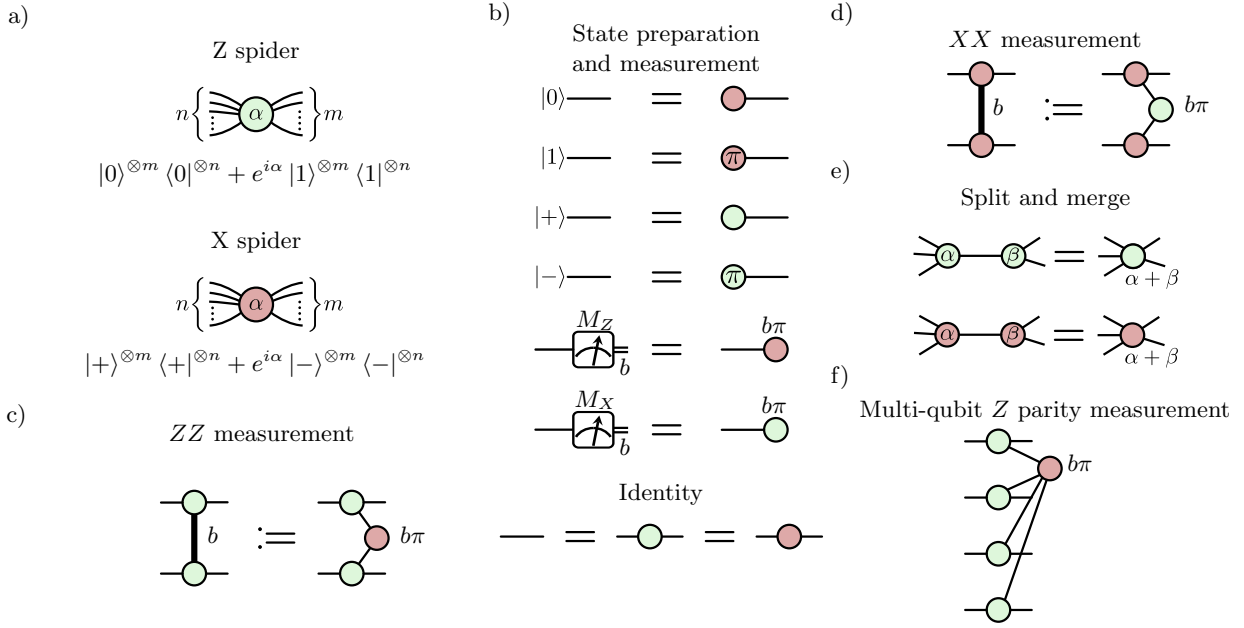}
    \caption{
    a) Z- and X-spiders: The basic building blocks for the ZX-calculus framework.
    b) Representing basic quantum operations using the ZX-calculus.
    c+d) We define an abbreviated symbol for the pairwise parity measurement of $XX$ or $ZZ$ observables. By dropping the measurement result, we can write two-qubit measurements as two same-color spiders.
    e) Basic split and merge relations on ZX-calculus nodes.
    f) A representation of a multiqubit parity measurement.
    Dropping the measurement outcome port in these circuits corresponds to reinterpreting a Pauli measurement as a projection.
    We refer to Ref.~\cite{Backens_2014ZX,Bombin_2024} for a rigorous introduction to ZX-calculus.
    }
    \label{fig:ZX-calculus}
\end{figure*}

The ZX-calculus is a useful tool to visualize and reason about quantum circuits. For a rigorous introduction, we refer the reader to Refs. \cite{Backens_2014ZX, Bombin_2024,van2020zx}. Here, we will only present identities which are useful for the rest of the paper.

We start by introducing the Z-spider and the X-spider nodes, presented in Fig.~\ref{fig:ZX-calculus}a. Each spider has $n$ input states and $m$ outputs and a phase $\alpha$. These spiders are equivalent to the operator 
$$\ket{+}^{\otimes m}\bra{+}^{\otimes n} + e^{i\alpha}\ket{-}^{\otimes m}\bra{-}^{\otimes n}$$
for an X-spider and 
$$\ket{0}^{\otimes m}\bra{0}^{\otimes n} + e^{i\alpha} \ket{1}^{\otimes m}\bra{1}^{\otimes n}$$
for a Z-spider. 

We can use those spiders to represent basic circuit operations such as initializations and measurement, as presented in Fig.~\ref{fig:ZX-calculus}b.

By connecting these spiders, we can represent the projection operator of a $k$-body measurement. This is illustrated in Figs.~\ref{fig:ZX-calculus}c and d for $k=2$, and can be generalized to any $k$ (for example, $k=4$ in Fig.~\ref{fig:ZX-calculus}f). We also define a notation to represent a 2-body pairwise measurement using bold edges, where $b\in\{0,1\}$ represent the classical measurement outcome.
This notation builds on the fact that when the measurement outcome is $0$, we can drop the central node in the measurement, as it is equivalent to the identity.

Those spiders can be split and merged with other same-type spiders as presented in Fig.~\ref{fig:ZX-calculus}e. 

\begin{figure}[t]
    \centering
    \def\svgwidth{0.6\linewidth}
    {
\begingroup%
  \makeatletter%
  \providecommand\color[2][]{%
    \errmessage{(Inkscape) Color is used for the text in Inkscape, but the package 'color.sty' is not loaded}%
    \renewcommand\color[2][]{}%
  }%
  \providecommand\transparent[1]{%
    \errmessage{(Inkscape) Transparency is used (non-zero) for the text in Inkscape, but the package 'transparent.sty' is not loaded}%
    \renewcommand\transparent[1]{}%
  }%
  \providecommand\rotatebox[2]{#2}%
  \newcommand*\fsize{\dimexpr\f@size pt\relax}%
  \newcommand*\lineheight[1]{\fontsize{\fsize}{#1\fsize}\selectfont}%
  \ifx\svgwidth\undefined%
    \setlength{\unitlength}{185.45886423bp}%
    \ifx\svgscale\undefined%
      \relax%
    \else%
      \setlength{\unitlength}{\unitlength * \real{\svgscale}}%
    \fi%
  \else%
    \setlength{\unitlength}{\svgwidth}%
  \fi%
  \global\let\svgwidth\undefined%
  \global\let\svgscale\undefined%
  \makeatother%
  \begin{picture}(1,1.65100882)%
    \lineheight{1}%
    \setlength\tabcolsep{0pt}%
    \put(0,0){\includegraphics[width=\unitlength,page=1]{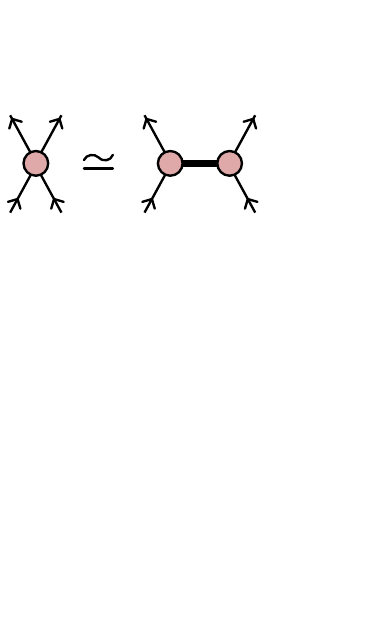}}%
    \put(0.05129507,0.45868593){\color[rgb]{0,0,0}\makebox(0,0)[lt]{\lineheight{1.25}\smash{\begin{tabular}[t]{l}Weight 8:\end{tabular}}}}%
    \put(0,0){\includegraphics[width=\unitlength,page=2]{decomposing-zx.pdf}}%
    \put(0.05100424,1.42116363){\color[rgb]{0,0,0}\makebox(0,0)[lt]{\lineheight{1.25}\smash{\begin{tabular}[t]{l}Weight 4:\end{tabular}}}}%
    \put(0.05115607,0.96823303){\color[rgb]{0,0,0}\makebox(0,0)[lt]{\lineheight{1.25}\smash{\begin{tabular}[t]{l}Weight 6:\end{tabular}}}}%
    \put(0.50657155,1.61168076){\color[rgb]{0,0,0}\makebox(0,0)[t]{\lineheight{1.25}\smash{\begin{tabular}[t]{c}Decomposing spiders into \\pairwise measurements\end{tabular}}}}%
  \end{picture}%
\endgroup%
}
    \caption{
    The interpretation of a $2n$-legged spider as an $n\rightarrow n$ operator is that of a partial projection onto a 2-dimensional subspace.
    The two-qubit pairwise parity measurements implement the target partial projection for one outcome configuration and the orthogonal projections for the others.
    Here we present different ways to factor a four-, six- and eight-spider into pairwise measurements. 
    The decomposition of the 8-spider intentionally includes one more measurement than necessary, introducing a local detector, a redundancy among the four measurement outcomes.
    Here, the symbol $\simeq$ denotes that equality up to a Pauli-frame, once a combination of measurement outcomes is fixed.
    }
    \label{fig:decompose-spiders}
\end{figure}

By combining those identities above, we can see that a 4-spider, 6-spider and 8-spider can be compiled into circuits with only pairwise measurement gates as presented in Fig.~\ref{fig:decompose-spiders}, up to a Pauli-frame.
The main property being preserved by this compilation is that the external stabilizers associated with the factorized gadget are the same as they would be for a single spider (up to signs which may depend on the measurement outcomes).
This allows preserving the original detectors through the compilation into a Floquet circuit.
These detectors are illustrated by closed Pauli webs which have no external support other than on the measurement outcome ports \cite{Bombin_2024} (See Sec~\ref{sec:detectors}).
Similarly, the correlator surfaces for logical operators will retain a similar structure.

As presented in Ref.~\cite{Bombin_2024}, a stabilizer code syndrome extraction circuit can be represented in the ZX-calculus language as a stack of Z-stabilizer-measurement and X-stabilizer-measurement layers.
This representation simplifies multiple operations, such as ancilla initialization, two-qubit entangling gates, and ancilla measurement into a single stabilizer measurement spider. Therefore, it can be interpreted as a ``phenomenological'' ZX-network representation.

While this simplification glosses over some important fault-tolerance features of the original circuit, such as its susceptibility to hook errors\footnote{These are single qubit errors on ancilla qubits which are equivalent to errors on more than one data qubit.}, it does reproduce many of the effective long range properties.
Similarly, the factorization of the spiders into pairwise Floquet measurements does not strictly preserve the fault-distance of the phenomenological ZX-network.
Given that we use the EM3 circuit-level noise model~\cite{Gidney_2021Honeycomb}
for the target circuit rather than the phenomenological per-edge noise model which is implicit in the ZX-network, it can be expected that the circuit-level distance would decrease.

The redundant measurement introduced in the decomposition of the weight 8 spider in Fig.~\ref{fig:decompose-spiders} introduces one additional internal detector, which is the joint parity of all four measurements involved (See~\ref{sec:detectors}). 
This detector will flag additional elementary circuit faults, which can increase the distance of the code~\cite{nickerson2018measurementbasedfaulttolerance}.

\subsection{Abelian two-block group algebra codes}

We start by a formal definition of Abelian 2BGA codes, and continue with a connection between those codes and geometrically-local codes within a $w{-}2$-dimensional space.

\subsubsection{Formal Definition} \label{sec:2bga-def}

Let $G = \{g_1, ... , g_\ell\}$ be a finite Abelian group of order $\ell$, 
and consider the group algebra $\mathbb{F}_2[G]$ of all formal linear combinations of elements of $G$ with coefficient in $\mathbb{F}_2$.
For $g\in G$, consider the permutation matrix $\mathbb{B}(g) \in \mathbb{M}_\ell(\mathbb{F}_2)$ which is defined by $\mathbb{B}(g)_{i,j}=1$ if and only if $g_i = g g_j$ (and 0 otherwise). This matrix describes the action of $g$ on $G$.

An Abelian 2BGA code is a CSS stabilizer code defined by the two binary matrices $H_X = [A|B]$ and $H_Z = [B^T|A^T]$, where $A = \sum_{g\in G} a_g \mathbb{B}(g)$ and $B = \sum_{g\in G} b_g \mathbb{B}(g)$, with $a_g, b_g \in \mathbb{F}_2$.
The number of qubits is $n=2 \ell$.
The code is defined by a set of stabilizer generators of weight $w = \textsf{wt}(a_g) + \textsf{wt}(b_g)$, where $\textsf{wt}$ denotes the Hamming weight (i.e. number of non-zero entries) of its argument.
This parameter $w$ corresponds to the weight referenced in the subsequent sections.

This code is a generalization of bicycle codes~\cite{kovalev2013bicycle_codes}, which appear when taking the group to be cyclic $G=\mathbb{Z}_n$, and also a generalization of bivariate bicycle codes~\cite{Bravyi_2024BB_codes}, which appear when taking the group to be a product of two cyclic groups $G=\mathbb{Z}_\ell \otimes \mathbb{Z}_m$.

\begin{figure*}
    \centering
    \def\svgwidth{\linewidth}
    {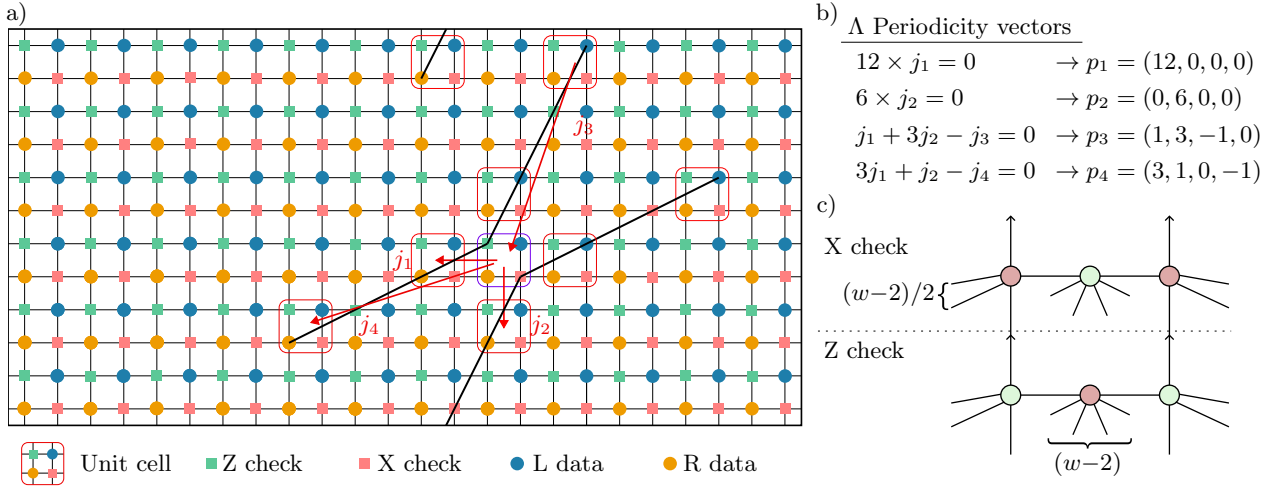}
    \caption{a) Example of converting the ``Gross code"~\cite{Bravyi_2024BB_codes} into a $D$-dimensional lattice. In this case, $D=w-2=4$, thus there are four basis vectors $j_1, \dots, j_4$. 
    The figure displays the code's Tanner graph; for visual clarity, only four long-range connections are shown. 
    One cell is highlighted in purple and its eight neighboring cells are highlighted in red. Four unit vectors are indicated by red arrows.
    b) The periodicity vectors $p_1,\dots, p_4$ of the generating sublattice $\Lambda$. 
    These vectors are associated to combinations of unit vectors which are identified as summing to zero.
    c) The basic lattice cell represented in the ZX-calculus formalism. 
    Vertical lines represent timelike connections while non-vertical lines represent spacelike connections.
    Each stabilizer spider have ${w}$ connections, and represents a $w$-wise parity measurement as presented in Fig.~\ref{fig:ZX-calculus}f.
    Each data spider have $\frac{w}{2}{+}2$ connections.
    While the example in this figure is a weight-6 code, in this manuscript we focus on weight-8 codes.
    }
    \label{fig:2bga}
\end{figure*}

\subsubsection{Abelian 2BGA Codes as Local Codes in a $(w{-}2)$-Dimensional Space} \label{sec:dimensional-2bga}
Abelian 2BGA codes can be modeled as local codes within a $D$-dimensional space (See Ref.~\cite{arnault2025_2bga_boundaries} for a formal proof), where $D=w-2$ and $w$ is the weight of the code stabilizers. The code's structure is generated from a specific sublattice of $\mathbb{Z}^D$, denoted as $\Lambda$, which defines the periodic boundary conditions. 
The resulting space is therefore a $D$-dimensional torus, formally described as the quotient space $\mathbb{Z}^D / \Lambda$.

We associate each point in this quotient space $\mathbb{Z}^D / \Lambda$ with a unit cell. 
Each cell is connected with their neighbors along $D$ directions, defined by a set of unit vectors $\{j_i\}_{i\in\{1,\dots,D\}}$.

To illustrate this geometrical framework, we consider the Gross code~\cite{Bravyi_2024BB_codes} as an example, which is a weight-6 code ($w=6$ and thus $D=4$). 
In Fig.~\ref{fig:2bga}a, one unit cell is outlined in purple, and its $2D=8$ neighboring cells are outlined in red. 
Each unit cell composed of two data qubits (blue and orange circles) and two stabilizer generators (green and red squares).
This structure generalizes to represent the 2-chain complex for other Abelian 2BGA codes, with the squares representing the checks (0-cells and 2-cell respectively).
The periodicity of this sublattice $\Lambda$ is defined by the connections of the stabilizer code, as shown in Fig.~\ref{fig:2bga}b. 
These vectors represent translations that map the lattice onto itself.

Those periodicity vectors can be classified into two types. 
The first type is axis-aligned; for example, $p_1 = (12,0,0,0)$ signifies that a translation of 12 steps along the $j_1$ direction returns to an equivalent lattice position ($12j_1=0$). 
The second type is non-axial, arising from linear combinations of unit vectors. 
As seen in the figure, the relationship $j_1 + 3j_2 = j_3$ (or $j_1 + 3j_2 - j_3 = 0$) holds, which corresponds to the non-axial periodicity vector $p_3 = (1,3,-1,0)$.

When there is only one axis-aligned periodicity vector, the resulting code is a bicycle code~\cite{kovalev2013bicycle_codes}.
For a sublattice with two axis-aligned periodicity vectors, the resulting code is a bivariate bicycle code~\cite{Bravyi_2024BB_codes}. 
Other orders can also be achieved with more (or less) axis-aligned periodicity vectors~\cite{voss2025multivariatebicyclecodes_trivariate}.

While this example illustrates the concept using a weight-6 code, this paper will focus on weight-8 ($D=6$) codes, as their foliated ZX-network consists only of even-weight spiders (Section \ref{sec:stabilizer-weight-constraint}).

The number of unit-cells for the 
code contained in this space is given by the determinant of the periodicity matrix:
\begin{align}
    \det \Lambda :=
    \det     \left|
    \begin{array}{c }
        p_1 \\
        p_2 \\
        \vdots \\
        p_D
    \end{array}
    \right| \;,
\end{align}
with each unit cell having two data qubits, one $X$-type stabilizer and one $Z$-type stabilizer associated with it.

The unit cell for a normalized syndrome extraction circuit (SEC), can also be represented in the ZX-calculus formalism~\cite{Bombin_2024}, as shown in Fig.~\ref{fig:2bga}c. In this depiction, the cell consists of four data spiders (two for each data qubit) and two auxiliary stabilizer spiders (one for each stabilizer). The connectivity of the spiders reflects the code's parameters: each stabilizer spider has $w$ connections, representing a weight-$w$ measurement, while each data spider has $w/2+2$ connections.
The syndrome extraction cycle has an additional direction $j_0$ which is not compacted by the code periodicity and is interpreted as time. 
In the traditional circuit model, data qubits' worldlines enter and exit the SEC unit cell in this direction. 
We define the unit vector in this dimension as a \emph{half}-cell step, which corresponds to the progression from a Z-stabilizer measurement to its associated X-stabilizer measurement.

\section{Floquet code from Abelian 2BGA codes}

Floquet codes are periodically driven syndrome extraction circuits, wherein measurements induce a dynamic instantaneous stabilizer group (ISG).
While the structure of the ISG is periodic and predetermined, the individual measurements driving the circuit need not be deterministic, leading to dynamically determined signs for the ISG generators.

In this way, Floquet codes generalize the traditional SECs for stabilizer codes and subsystem codes.
Similarly to subsystem SECs and contrary to stabilizer SECs, individual measurement outcomes in Floquet SECs are allowed to be random.
Traditional SECs for subsystem codes associate all deterministic outcome combinations to the stabilizer subgroup. In the absence of errors, this is a part of the ISG with pre-determined sign.
In contrast, in Floquet, no part of the induced ISGs is required to have pre-determined sign.
Thus, while deterministic combinations of measurement outcomes (a.k.a. detectors) still exist and form the basis of error diagnosis, these no longer need to be associated to a persistent stabilizer and may require combining measurement outcomes from multiple time steps.

\subsection{Stabilizer Weight Constraints} \label{sec:stabilizer-weight-constraint}

\begin{figure}
    \centering
    \def\svgwidth{\linewidth}
\begingroup%
  \makeatletter%
  \providecommand\color[2][]{%
    \errmessage{(Inkscape) Color is used for the text in Inkscape, but the package 'color.sty' is not loaded}%
    \renewcommand\color[2][]{}%
  }%
  \providecommand\transparent[1]{%
    \errmessage{(Inkscape) Transparency is used (non-zero) for the text in Inkscape, but the package 'transparent.sty' is not loaded}%
    \renewcommand\transparent[1]{}%
  }%
  \providecommand\rotatebox[2]{#2}%
  \newcommand*\fsize{\dimexpr\f@size pt\relax}%
  \newcommand*\lineheight[1]{\fontsize{\fsize}{#1\fsize}\selectfont}%
  \ifx\svgwidth\undefined%
    \setlength{\unitlength}{110.97127839bp}%
    \ifx\svgscale\undefined%
      \relax%
    \else%
      \setlength{\unitlength}{\unitlength * \real{\svgscale}}%
    \fi%
  \else%
    \setlength{\unitlength}{\svgwidth}%
  \fi%
  \global\let\svgwidth\undefined%
  \global\let\svgscale\undefined%
  \makeatother%
  \begin{picture}(1,0.91064506)%
    \lineheight{1}%
    \setlength\tabcolsep{0pt}%
    \put(0,0){\includegraphics[width=\unitlength,page=1]{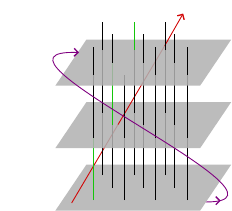}}%
    \put(0.79326767,0.88003986){\color[rgb]{0.82352941,0.02745098,0.02745098}\makebox(0,0)[lt]{\lineheight{1.25}\smash{\begin{tabular}[t]{l}Time\end{tabular}}}}%
    \put(0.57759455,0.90564287){\color[rgb]{0.08235294,0.77254902,0.04313725}\makebox(0,0)[t]{\lineheight{1.25}\smash{\begin{tabular}[t]{c}Qubit\\worldline\end{tabular}}}}%
    \put(0.10371878,0.67160638){\color[rgb]{0.50196078,0,0.50196078}\makebox(0,0)[t]{\lineheight{1.25}\smash{\begin{tabular}[t]{c}Periodicity\\vector\end{tabular}}}}%
    \put(0,0){\includegraphics[width=\unitlength,page=2]{time_rotation.pdf}}%
  \end{picture}%
\endgroup%

    \caption{
    Geometry of the spacetime lattice. 
    The green trajectory, resembling the path of a stairway, indicates the worldline of a physical qubit in a Stairway code. 
    Every step has a positive projection onto the time vector. Thus, the qubit consistently advances in time. 
    The $D$-dimensional spatial layers containing original stabilizer checks are shown as grey planes. Notably, the periodicity vector is chosen to be orthogonal to the time vector, preventing closed timelike curves and keeping the time coordinate well-defined.
    }
    \label{fig:schematic-3d}
\end{figure}

Unlike other constructions of Abelian 2BGA codes, such as the bivariate bicycle construction which uses weight-6 stabilizers~\cite{Bravyi_2024BB_codes}, our method requires weight-8 stabilizers for the underlying code. This section explains the origin of this constraint.

Our strategy for converting Abelian 2BGA codes into Floquet codes involves reinterpreting the simplified ZX-tensor network representation of the standard syndrome extraction circuit of the code as a new set of qubit ``worldlines'' and pairwise measurements. 
In order to admit such an interpretation, every spider node in the simplified ZX-diagram is required to have an even number of legs.
This property is required to allow pairing the legs of each spider node into corresponding ``input'' and ``output'' legs, which admit an interpretation as qubit worldlines.

The Abelian 2BGA codes we consider are CSS codes where all stabilizers have a uniform weight $w$, and each data qubit couples to $w/2$ X-stabilizers and $w/2$ Z-stabilizers. This structure translates to two types of nodes in the ZX-diagram (see Fig.~\ref{fig:2bga}c):
\begin{itemize}
\item \textbf{Stabilizer spiders:} These correspond to the stabilizer measurements and each have $w$ legs (connections).
\item \textbf{Data spiders:} These represent the data qubits and have $w/2$ spatial connections plus two temporal connections (linking to the previous and next time step), for a total of $w/2+2$ legs.
\end{itemize}

Since we require the number of legs of both spider types to be even (i.e. same number of outputs as inputs), this imposes the constraint:
\begin{align}
    w \equiv 0 \pmod{4} \;.
\end{align}

Lower stabilizer weights $w$ are desirable, as they simplify fabrication through reduced hardware connectivity requirements.
Furthermore, we expect lower-weight detectors to be more reliable at diagnosing errors and provide more actionable information to the decoder.
The lowest weight permitted by our constraint is $w=4$ and corresponds to the toric code \cite{Kitaev2003} and its ``Floquetification'' \cite{Bombin_2024}, either of which can be arranged in a geometrically local way on a 2D torus but supports only $k=2$ logical qubits.
In order to keep the weight of detectors as small as possible, we focus on the next lowest compatible value of $w=8$ for the rest of the paper.
While higher weights such as $w=12$ could in principle yield codes with higher rates and distances, the increased detector weight and connectivity cost make $w=8$ the most practical starting point.

\subsection{Reinterpreting time in a ZX-network} \label{sec:zx-time-rotation}

We have now established a representation of Abelian 2BGA codes as foliated ZX-networks. We also introduced an efficient way of compiling the network into pairwise measurements (Fig.~\ref{fig:decompose-spiders}).
Yet, these components alone are insufficient; the majority of connections in the foliated network are spatial, and thus the ``virtual" qubits that they represent do not translate to a circuit model.

To address this, we introduce a key component of our construction: a rotated time direction. 
We define a time covector that is non-orthogonal to any of the lattice's generating unit vectors,
permitting reinterpretation of the graph's spatial edges as physical qubit worldlines advancing through time (Fig.~\ref{fig:schematic-3d}).

As explained in Section~\ref{sec:dimensional-2bga}, the ZX-calculus graph of the syndrome extraction circuit for a weight-$w$ code can be arranged as a lattice of unit cells in a $(w{-}1)$-dimensional space-time. 
This lattice is composed of two alternating types of half-cells: a \textbf{Z-check half-cell}, where a central X-spider measures a Z-product, and an \textbf{X-check half-cell}, where the roles are swapped. Half-a-step in the original time-direction, ${j}_0$, moves from a Z-type half-cell to an X-type half-cell (as the vertical connections in Fig. \ref{fig:2bga}c).

A vector in this space can be written in the basis of $(j_0, j_1, j_2, \dots j_{D})$ where the first coordinate denotes the original circuit's time direction, and the remaining coordinates correspond to the spatial dimensions. 
We mark the unit vectors as $\hat{j}_i$ with $i\in\{0,...,D\}$.

\begin{figure*}
    \centering
    \def\svgwidth{\linewidth}
    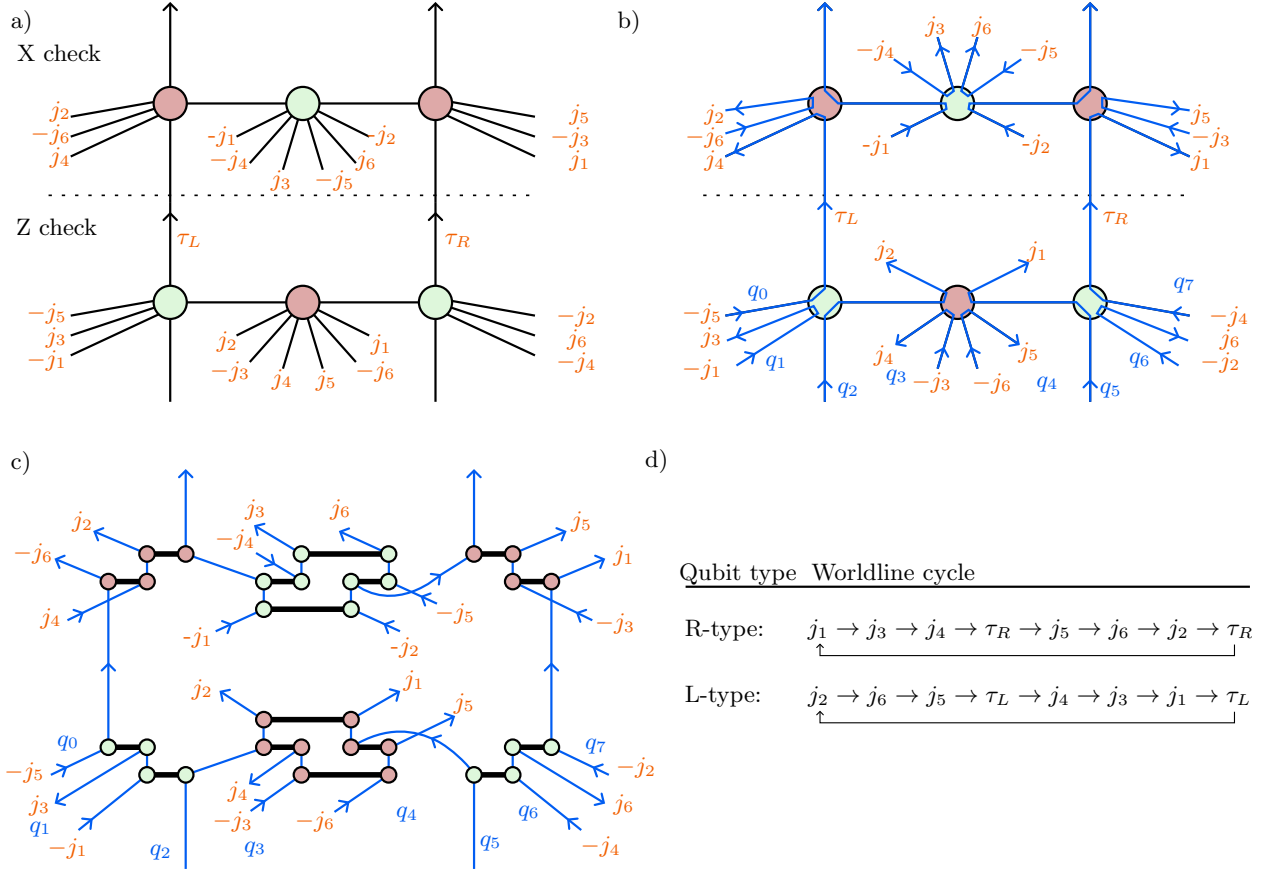
    \caption{The process of decomposing the weight-8 Abelian 2BGA unit-cell into pairwise measurements.
    a) The unit cell and the directions to neighboring unit cells. Originally, each unit cell is composed of two data qubits and two stabilizer measurements. 
    We divide the unit cell into two parts, which we call half-cells, that represent the X check and Z check.
    b) We define stairway qubit worldlines (blue) by tracing through the contracted edges of the ZX network. 
    The qubit worldlines progress in the direction of the unit vectors, thus coming from a half-cell with a lower timestep and leaving to a half-cell with a higher timestep.
    c) Using the rules from Fig.~\ref{fig:ZX-calculus}, we factor the high connectivity spiders into pairwise measurements. 
    This step introduces 3 sub-steps in each half-cell, as some qubits participate in three gates. 
    For simplicity, we treat each half-cell as point-like, neglecting the deviations from the unit vectors caused by internal displacements.
    d) By tracing the worldlines, we track a cycle spanning 8 timesteps. We present the two possible trajectories a worldline can follow.
    }
    \label{fig:basic-cell}
\end{figure*}

By defining
\begin{align}
    t=(2,1,\dots,1) \;,
\end{align}
we effectively ``rotate" the time direction. 
Under this choice, each spatial direction also advances time: when a qubit moves along any direction $\hat{j}_i$, it actually moves forward in time. 
Furthermore, moving in the direction of time still advances time. 
In particular, moving half a unit vector in the time direction, which we define as $\tau = \frac{1}{2} j_0$, is equivalent to one timestep, and moving a unit vector in the spatial direction is also equivalent to one timestep.
A schematic representation is given in Fig.~\ref{fig:schematic-3d}.
Here, we choose a small rotation angle. Larger rotations might yield other interesting results.

Therefore, every space-time unit cell can be divided into two half-cells (see Fig.~\ref{fig:basic-cell}a). 
Each half-cell has $D{+}1$ neighboring half-cells that are ``one time-step ahead'' as well as $D{+}1$ neighboring half-cells which are ``one time-step behind''. 
The connections to those neighboring half-cells correspond to qubits coming from another half-cell which are earlier in time and then leaving into half-cells which are later in time.
Notably, qubits that enter a cell can exit in other directions that are not the original circuit time direction $\hat{j}_0$.

To calculate the time for a half-cell, we take the inner product of the cell location $l$ with the time vector $l \cdot t$, and add $1$ for an X-type half-cell. We mark this transformation between the lower half of a unit cell (Z-check half-cell) to the upper half (X-check half-cell) as $\tau$, as can be seen in Fig.~\ref{fig:basic-cell}a.

When rotating the time direction, our boundary conditions also need to change. 
Our space is now $(D{+}1)$-dimensional, and its periodicity vectors should be orthogonal to the time direction $p'_i \cdot t = 0$.
Thus, each code in the Stairway code family is defined by a different set of periodicity vectors.
Importantly, this choice of boundary conditions determines the number of logical qubits the code can hold, and the distance they exhibit.
However, similar to the case of bivariate bicycle codes, this dependence does not have a simple closed-form expression.
The new periodicity vectors are non-spatial, which is a form of time-vortex~\cite{kishony2025increasingdistancetopologicalcodes}.

We denote the new lattice periodicity matrix, composed of $D$ vectors $p'_i$, with $D+1$ components each as $\Lambda'$.

\subsection{Decomposing the ZX-network} \label{sec:zx-decomposition}

After taking a new interpretation for the direction in which time progresses in the ZX-network, we are still left with the freedom of matching the inputs and outputs of each spider, specifying an interpretation as qubit worldlines.

There are multiple consistent ways to perform this selection, corresponding to different valid worldline configurations. 
In Fig.~\ref{fig:basic-cell}b, we present one particular choice, which serves as a representative example for our construction by highlighting the qubit worldlines in blue. 
A review of the different possible ways to perform this decomposition is given in the Methods Section.
Using the rules from Fig.~\ref{fig:decompose-spiders}, we can split each spider in the cell into a series of operations, effectively defining our circuit. This is presented in Fig.~\ref{fig:basic-cell}c.

By tracing the qubit worldlines and registering their interactions, the full circuit structure emerges. 
Each qubit follows a periodic schedule, as illustrated in Fig.~\ref{fig:basic-cell}d. 
This tracing procedure can be carried out by tracking each qubit within the cell. As an illustration, a qubit that enters a Z-type check from direction $j_6$ exits the cell along direction $j_5$.

This schedule has a period of 8 time steps (equivalent to 24 sub-steps, as each half-cell includes 3 sub-steps). 
This is significantly longer than other Floquet codes, such as the standard Honeycomb code (3 sub-steps) or the CSS Honeycomb code (6 sub-steps). While such an extended period might be expected to affect the logical error rate due to error accumulation, we demonstrate later that the logical performance remains robust.

We refer to this period as one quantum error correction round. 
Over the course of one round, each qubit effectively undergoes a constant coordinate displacement $\vec{\delta}$ within the 7-dimensional space. 
This means that the circuit perfectly repeats itself after 8 time steps, ensuring that each qubit interacts with a closed set of other qubits.
Specifically, we observe that each qubit interacts with a fixed set of $10$ qubits.
If we wish to interpret the circuit as composed of static physical qubits, we may do so by defining spatial covectors $\hat{x}$ and $\hat{y}$, which are orthogonal to the coordinate displacement in the SEC cycle (i.e. $\hat{x}\cdot\vec{\delta}=0$ and $\hat{y}\cdot\vec{\delta}=0$).
This allows us to use $\hat{x}$ and $\hat{y}$ to associate spatial coordinates to qubit worldlines (up to small straightenable displacements which are periodic in the cycle).

Like a stairway, which always goes up but also keeps some component in a varying orthogonal direction, the qubit worldlines in stairway codes always progress along a chosen temporal direction while cycling their orthogonal components in the embedding lattice, granting our code its name.

The total number of physical qubits in the code can be calculated from the number of half-cells sharing a time step, multiplied by the number of qubits participating in each half-cell, $w$($=8$ in our case). 
This can be calculated to be:
\begin{align} \label{eq:number-of-qubits}
     w \det \Lambda' /{\Vert t \Vert}_1 \;,
\end{align}
to further illustrate, Table \ref{table:codes_generators} lists the sublattice generator $\Lambda'$ for three of the codes given in Table \ref{table:codes_intro}.

\begin{table}[t]
    \centering
    \begin{tabular}{c|c}
        $[[n,k,d]]$ & 
        \begin{tabular}{c} 
            $\Lambda'$ \\ 
            $\left(\begin{array}{*{7}{wc{1.3em}}} j_0 & j_1 & j_2 & j_3 & j_4 & j_5 & j_6 \end{array}\right)$ 
        \end{tabular}
        \\
        \hline
         $[[192,16,4]]$
         &
        $\left(
        \begin{array}{*{7}{wc{1.3em}}}
            -2 & 4 & 0 & 0 & 0 & 0 & 0 \\
            -3 & 0 & 6 & 0 & 0 & 0 & 0 \\
            -2 & 0 & 5 & -1 & 0 & 0 & 0 \\
            -1 & 3 & 0 & 0 & -1 & 0 & 0 \\
            -3 & 3 & 4 & 0 & 0 & -1 & 0 \\
            -2 & 3 & 2 & 0 & 0 & 0 & -1  
        \end{array}
        \right)$
         \\
         \hline
          $[[288,14,\le10]]$
          &
         $\left(
         \begin{array}{*{7}{wc{1.3em}}}
              -3 & 6 & 0 & 0 & 0 & 0 & 0 \\
              -3 & 0 & 6 & 0 & 0 & 0 & 0 \\
              -2 & 3 & 2 & -1 & 0 & 0 & 0 \\
              -4 & 5 & 4 & 0 & -1 & 0 & 0 \\
              -2 & 4 & 1 & 0 & 0 & -1 & 0 \\
              -3 & 4 & 3 & 0 & 0 & 0 & -1
                     \end{array}
         \right)$
         \\
        \hline
        $[[576,14,\le 20]]$
        &
        $\left(
        \begin{array}{*{7}{wc{1.3em}}}
            -3 & 6 & 0 & 0 & 0 & 0 & 0 \\
            -6 & 0 & 12 & 0 & 0 & 0 & 0 \\
            -4 & 4 & 5 & -1 & 0 & 0 & 0 \\
            -1 & 1 & 2 & 0 & -1 & 0 & 0 \\
            -3 & 2 & 5 & 0 & 0 & -1 & 0 \\
            -5 & 1 & 10 & 0 & 0 & 0 & -1          
        \end{array}
        \right)$
    \end{tabular}
    \caption{Code parameters $[[n,k,d]]$ and the sublattice generator $\Lambda'$ which produced the code. Note that all periodicity vectors are orthogonal to the time covector $t\cdot p_i=0$ (see Sec.~\ref{sec:zx-time-rotation}).
    }
    \label{table:codes_generators}
\end{table}

\subsection{Generation of the Floquet Schedule}

While the code is theoretically defined by a ZX-calculus graph embedded in a $(w{-}1)$-dimensional space, the practical implementation requires a deterministic schedule $\mathcal{S}$ that dictates which qubit pairs interact at each time step.

We formalize this process in Algorithm \ref{alg:stairway_schedule}, which systematically maps the high-level geometric parameters—specifically the periodicity matrix $\Lambda'$ and the time covector $t$—into a linear circuit schedule. The algorithm iterates through the fundamental domain of the lattice, projecting each unit cell onto the time axis to establish a base timeline. It then applies the local decomposition template $\mathcal{T}_{decomp}$ (derived from the spider decomposition in Fig.~\ref{fig:basic-cell}c ) to generate the specific sequence of pairwise measurements\footnote{Still confused? Check out the supplementary scripts published in~\cite{jacoby2026data}.}.

\begin{algorithm}[t]
\caption{Stairway Code Schedule Generation}
\label{alg:stairway_schedule}
\begin{algorithmic}[1]
\Require Periodicity matrix $\Lambda'$, Time covector $t$, Local template $\mathcal{T}_{decomp}$, $T_{\min}$, $T_{\max}$.
\Ensure Schedule $\mathcal{S}$ mapping time-steps to pairwise measurements.

\State $\mathcal{S} \gets \text{defaultdict(list)}$
\State \PyComment{Get all coordinates $\vec{u}$ in lattice}
\State $U \gets \text{GenerateUnitCells}(\Lambda', t, T_{\min}, T_{\max} )$ 

\ForEach{cell coordinate $\vec{u} \in U$}
    \State \PyComment{Project spatial coord onto time axis}
    \State $T_{base} \gets \vec{u} \cdot t$ 
    
    \ForEach{half-cell type $h \in \{Z, X\}$}
        \State \PyComment{X-check is offset by 1 step relative to Z-check}
        \State $T_{start} \gets T_{base} + (1 \text{ if } h=X \text{ else } 0)$
        
        \ForEach{op $(\delta_\tau, q_a, q_b, P) \in \mathcal{T}_{decomp}[h]$}
            \State \PyComment{Apply sub-step offset}
            \State $T_{op} \gets T_{start} + \delta_\tau$ 
            
            \State \PyComment{Resolve physical IDs of qubits}
            \State $Q_1 \gets \text{ResolveID}(\vec{u}, h, q_a, \Lambda')$
            \State $Q_2 \gets \text{ResolveID}(\vec{u}, h, q_b, \Lambda')$
            
            \State $\mathcal{S}[T_{op}].\text{append}(\text{Measure}(Q_1, Q_2, P))$
        \EndFor
    \EndFor
\EndFor

\State \Return $\mathcal{S}$
\end{algorithmic}
\end{algorithm}

\section{Results}

\subsection{Distance} \label{sec:distance-results}

While code distance is a well-defined concept for static stabilizer codes, i.e. the lowest Hamming weight of an operator commuting with all the stabilizer group, for dynamical codes we must focus on the {\it{fault-distance}}~\cite{fu_Gottesman2025errorcorrectiondynamicalcodes}.
This is defined as the lowest weight combination of ``elementary faults'', which implements a logical operation while remaining undetected, but requires a reference definition of what ``elementary faults'' are. 

In this work, we consider two distances. 
The first is the \textit{embedded distance}, defined as the minimal distance of the embedded homological code when considering all timesteps~\cite{Hastings_2021}. 
In this code, at the timestep subsequent to a pairwise measurement between two qubits, we regard the two measured qubits as one effective qubit, as the measurement projects them into a two-dimensional state.
Appendix~\ref{sec:embedded-distance-for-CNOT} presents another perspective on embedded distance.

The second is the circuit-distance $d_\text{circ}$ -- how many physical errors can lead to an undetected logical error, defined for a specific noise model. 
In this work, we adapt the EM3 noise model (Entangling measurements) from Refs.~\cite{Gidney_2021Honeycomb, Higgott_2024hyperbolic, fahimniya2025hyperbolicfloquet}. 
This error rate is tailored for systems with native pairwise measurements. 
In the EM3 noise model, pairwise measurements are subject to an error with probability $p$. This error introduces a fully depolarizing channel on the check's qubits and flips the measurement outcome in half of these error events.

We can calculate the code distance by analyzing the circuit generated from a periodicity matrix $\Lambda'$ (see Sec.~\ref{sec:finding-distance}). We use this method to scan a generated pool of candidates. As the search narrows down to the most promising codes, we increase the computational resources used to verify the distance for each instance. Examples of the resulting codes are shown in Table~\ref{table:codes_intro}.

\begin{figure}
    \centering
    \includegraphics[width=\linewidth]{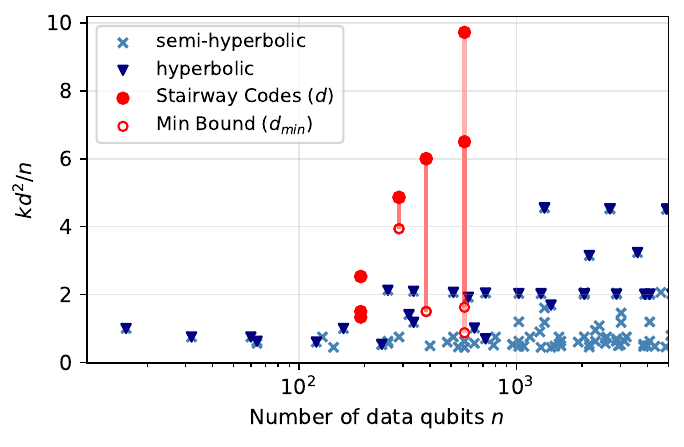}
    \caption{Performance comparison using the $kd^2/n$ metric relative to hyperbolic and semi-hyperbolic Floquet codes~\cite{Higgott_2024hyperbolic}. Vertical lines denote the uncertainty interval $[d_{\min}, d]$ for instances where the ILP solver did not converge to a single value. 
    }
    \label{fig:hyperbolic-metric-comparison}
\end{figure}

A comparison of the resulting code parameters, quantified by the $kd^2/n$ metric, is presented in Fig.~\ref{fig:hyperbolic-metric-comparison}.

\subsection{Logical Error rate}\label{sec:LER_results}

\begin{figure}
    \centering
    \includegraphics[width=\linewidth]{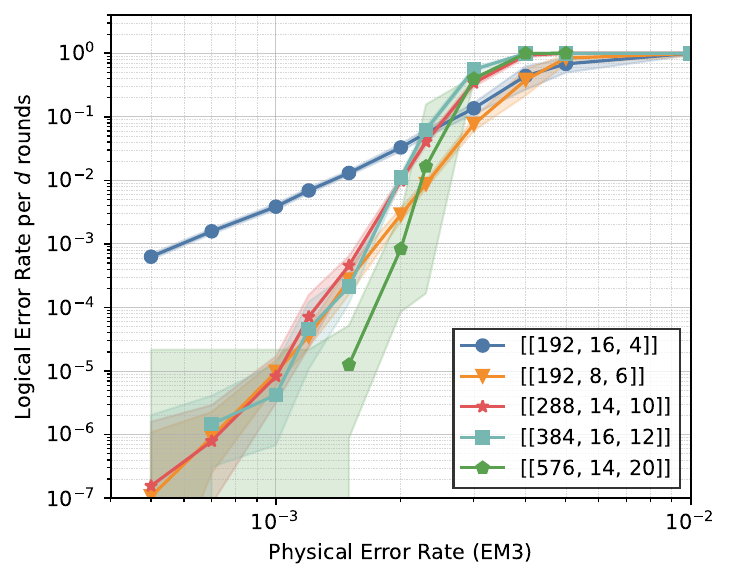}
    \caption{Logical error rate of Stairway codes obtained from d rounds of Monte-Carlo sampling. Due to decoding being computationally infeasible for the $[[576,14,\le 20]]$ code, we sampled it for 6 rounds and renormalized it to 20 (see Appendix \ref{sec:appndx-decoding-time}). Error bars are for 99\% confidence.
    }
    \label{fig:logical-error-rate}
\end{figure}

In this section, we benchmark the showcased Stairway codes using a Clifford simulation and compare them to hyperbolic Floquet codes~\cite{Higgott_2024hyperbolic} and to bivariate bicycle codes with syndrome extraction circuit compiled into pairwise measurements. We present the logical error rate of a $d$-round memory experiment. For these simulations, we again use the EM3 noise model~\cite{Gidney_2021Honeycomb}.
For decoding, we have used the tesseract decoder~\cite{beni2025tesseractdecoder}. We have tried using BP+OSD decoder~\cite{Roffe_2020bposd} as well, but encountered an inferior error rate.

The logical error rate of a memory experiment consisting of $d$ rounds is presented in Fig.~\ref{fig:logical-error-rate}. It is evident that the logical error rate suppression is significantly stronger than implied by the circuit-level distance given in Table~\ref{table:codes_intro}, a behavior attributed to the dominance of high-weight errors in the near-threshold regime~\cite{Beverland_2019}. 
The decoding of the $[[576,14,20]]$ for 20 rounds was too computationally expensive, thus we sampled it for 6 rounds and renormalized to 20 (Appendix \ref{sec:appndx-decoding-time}).

\begin{figure}[t]
    \centering
    \includegraphics[width=\linewidth]{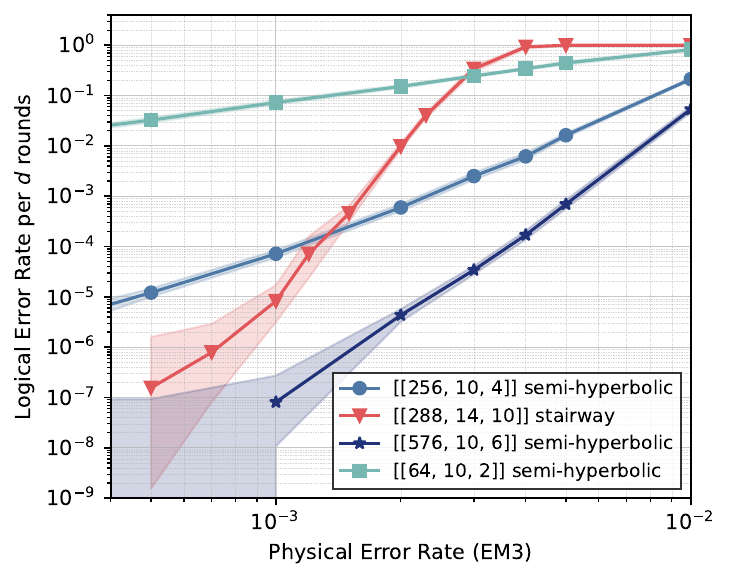}
    \caption{Logical error rate vs. physical error rate ($10$ rounds) for the semi-hyperbolic Floquet codes~\cite{Higgott_2024hyperbolic} and the $[[288,14,10]]$ stairway code. Although our code exhibits a lower threshold, it outperforms the lower-distance ($[[64,10,2]]$ and $[[256,10,4]]$) semi-hyperbolic codes in the low-error regime. We note that the circuit distance of the $[[288,14,10]]$ code is $\dcirc \le 7$; thus, we expect the scaling in the low-error regime to be only marginally superior to that of the distance-6 semi-hyperbolic code.
    }
    \label{fig:hyperbolic-comparison}
\end{figure}

To further benchmark our codes, we compare them to the semi-hyperbolic Floquet codes presented by \textit{Higgott et al.}~\cite{Higgott_2024hyperbolic}. 
While a direct comparison is challenging due to the differing number of logical qubits, we focus on a set of semi-hyperbolic codes with k=10 and compare them against the $[[288,14,\le10]]$ stairway code. 
We plot the logical error rates for these codes in Fig.~\ref{fig:hyperbolic-comparison}.

It can be seen that due to the lower threshold exhibited by the stairway codes, the semi-hyperbolic codes are superior in the high-error regime. However, the stairway code outperforms the lower-distance semi-hyperbolic codes as the physical error rate decreases. 
Specifically, for an error rate of $p=0.1\%$, the $[[288,14,10]]$ stairway code outperforms the $[[256,10,4]]$ semi-hyperbolic code while hosting more logical qubits (14 vs. 10), achieving a higher encoding rate with roughly the same block size (288 vs. 256).

Finally, we compare our performance against bivariate bicycle codes, specifically the $[[144,12,12]]$ Gross code, with syndrome extraction compiled into pairwise measurements as detailed in Section~\ref{sec:BB-codes}.

We introduce two compilation strategies. 
The first, which we term the ``long'' construction, compiles each pair of CNOTs independently. This preserves the circuit-level distance but introduces significant idling noise.

The second strategy, the ``short'' construction, adopts a compilation approach similar to that of Gidney et al.~\cite{Gidney_2023}. While this yields a compact syndrome extraction circuit with minimal idling, it reduces the circuit-level distance upper bound from $\dcirc \le 10$ (in the original CNOT-based construction) to $\dcirc \le 8$.

Both constructions require three ancillas per stabilizer, effectively doubling the total number of qubits compared to the original construction. 

\begin{figure}[t]
    \centering
    \includegraphics[width=\linewidth]{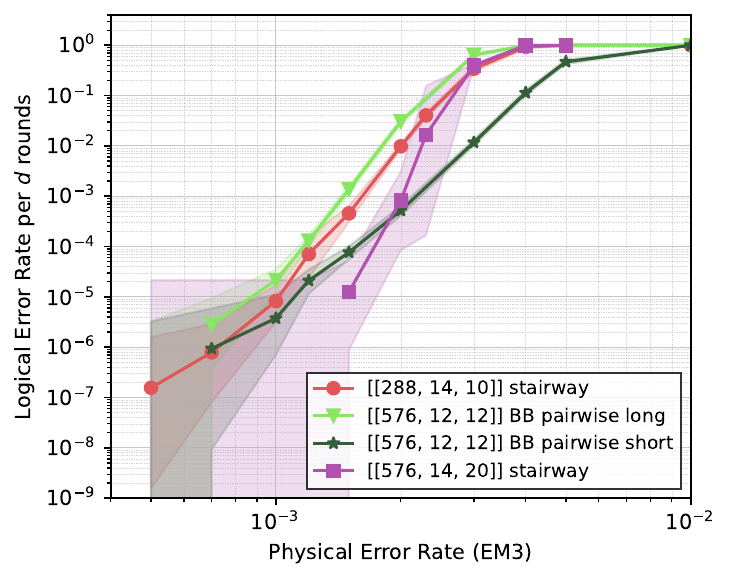}
    \caption{Logical error rate versus physical error rate ($d$ rounds) for two stairway codes and the Gross code. The Gross code is compiled into pairwise measurements using the two strategies described in Section~\ref{sec:BB-codes}, to which we refer as ``BB pairwise long" and ``BB pairwise short". In the label $[[n,k,d]]$, $n$ denotes the total number of qubits, including ancillas.}
    \label{fig:BB-comparison}
\end{figure}

The logical error rates for these benchmarks are presented in Fig.~\ref{fig:BB-comparison}, where it can be seen that the stairway codes outperform both Gross code constructions.

\section{Conclusion}

The Stairway codes family is a promising candidate for quantum error correction 
with native pairwise measurements.

This code family has a promising structure, as it builds on top of the well-studied two-block group algebra codes, specifically bivariate bicycle codes, which provide an efficient numerical framework for searching code parameters. 
While a full search is beyond current numerical methods, the examples we have found and examined, presented in Table~\ref{table:codes_intro}, already exhibit state-of-the-art code parameters.

This code family has been designed using a novel combination of methods. 
We have considered two-block group algebra codes, and their (foliated) syndrome extraction circuits as ZX-networks in a $(w{-}2)$-dimensional space.
By re-interpreting the direction of time on this representation, we have constructed different ``Floquetified'' quantum memory protocols. 
Finally, by decomposing the ZX-calculus spiders, we simplified the native gate-set to require only pairwise-measurements, introducing new detectors in the process.

In this article, we have focused on finite size code instances rather than asymptotic code families. For Abelian 2BGA codes embedded in 2D, the dependence of the number of logical qubits $ k $ and the code distance $ d $ will in general have an intricate dependence on the choice of boundary conditions \cite{Liang2025}. 
Although in our context, the $ D+1 $ syndrome extraction protocol is quotiented by a lattice $ \Lambda' $, we expect the scaling bounds for code distance $ d $ derived in Ref.~\cite{arnault2025_2bga_boundaries}, which assumes a geometrically local code on a lattice quotient $ \mathbb{Z}^D/\Lambda $, to apply. 
Technically, for the statement to hold, one would need to guarantee that the ISGs for the Stairway codes can be expressed in terms of geometrically local generators in $ D $ dimensions.

To present a logical error rate, we have used the Tesseract decoder~\cite{beni2025tesseractdecoder}.
This produces a state-of-the-art logical error rate, with the caveat of long decoding times. 
We suspect those times can be improved by considering other decoding strategies, such as BP+DTD~\cite{ott2025decisiontreedecoders-DTD} or other general qLDPC decoders~\cite{müller2025relayBP, ye2025beamsearchdecoderquantum}.

While we focused mainly on codes with lattice geometry that resembles a bivariate bicycle code, Stairway codes can Floquetify other Abelian 2BGA codes, such as bicycle codes~\cite{kovalev2013bicycle_codes} and trivariate bicycle codes~\cite{voss2025multivariatebicyclecodes_trivariate}. 
Those are left to be explored.

\section{Methods}

\subsection{From ZX-network to Floquet circuit}

As noted in Section~\ref{sec:zx-time-rotation}, there is no unique way to interpret the ZX-network obtained as a normal form of the Abelian 2BGA circuit model as a Floquet syndrome extraction protocol. 
Here, we outline the specific principles and methodology used to derive the transformation presented in this work. 
The process is divided into two distinct stages: identifying new qubit worldlines and decomposing high-degree spiders into pairwise measurements.

\subsubsection{Stage 1: Worldline Assignment}

The first stage involves identifying qubit worldlines within the graph by assigning an input direction and an output direction to each of the eight qubits in a half-cell. 
The specific configuration chosen for this work is depicted in Fig.~\ref{fig:basic-cell}b.

We guided this selection using the following criteria:
\begin{itemize}
\item \textbf{Symmetry:} We enforced symmetry between the left-type (L) and right-type (R) spiders to ensure a balanced code structure.
\item \textbf{Uniform Propagation:} We sought a configuration that creates a closed ``cycle" where all qubits undergo uniform displacement. 
In our chosen configuration, after eight time steps, every qubit advances by the exact displacement vector $(1,1,1,1,1,1,1)$, ensuring consistent movement across the lattice.
\end{itemize}

While not an essential consideration, we also note that the chosen qubit worldlines identification reproduces the construction in Ref.~\cite{Bombin_2024} when qubits $q_1,q_3,q_4,q_6$ are removed.

\subsubsection{Stage 2: Spider Decomposition}

In the second stage, we apply the ZX-calculus reduction rules (see Fig.~\ref{fig:ZX-calculus}e) to decompose the high-degree spiders (degree-6 data spiders and degree-8 stabilizer spiders) into sequences of pairwise measurements.

While there are many ways to perform this decomposition, we tested several different candidates and observed no notable difference between them, either numerically or analytically. We selected the configuration presented in Fig.~\ref{fig:basic-cell}c, as it is both symmetric and introduces only 3 sub-steps. 
A configuration with lower-depth can be achieved by removing one of the measurements decomposed from the 8-spider, at the cost of losing the corresponding detector.

\subsection{Identifying Detectors} \label{sec:detectors}

\begin{figure}
    \centering
    \centering
    \def\svgwidth{\linewidth}
    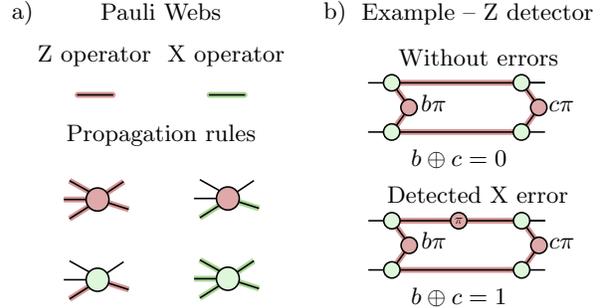
    \caption{Pauli webs are used to visualize the flow of operators on the ZX-network. We can use those flows to identify detectors.}
    \label{fig:pauli-webs}
\end{figure}

In fault-tolerant quantum computation, detectors are sets of measurements whose product produces a deterministic value in the absence of errors. 
For example, in stabilizer-based error correction, a detector is typically defined by the product of two consecutive measurements of the same stabilizer generator.

We use the \textit{Pauli web} overlay notations to analyze the propagation of operators through the ZX tensor network diagram.
Using the rules presented in Fig.~\ref{fig:pauli-webs}a, we can systematically identify these loops.

\begin{figure}
    \centering
    \centering
    \def\svgwidth{\linewidth}
    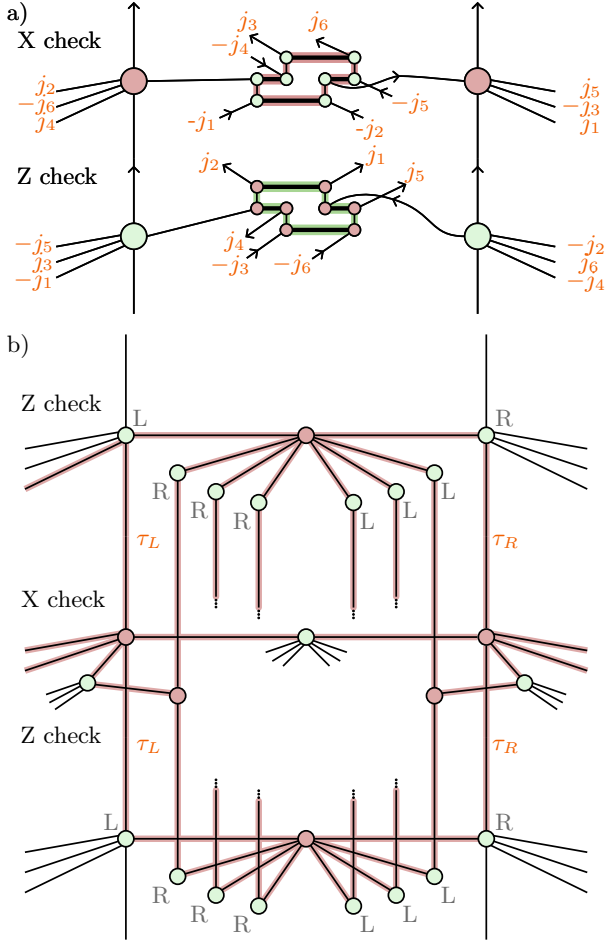
    \caption{Detectors in the stairway codes. (a) A localized detector resulting from the decomposition of a weight-8 spider into four pairwise measurements. 
    (b) A large detector, spanning 14 unit cells and comprising the parity of 32 pairwise measurement outcomes.
    The internal decomposition of weight 8 and 6 spiders, present in Fig \ref{fig:basic-cell}c is omitted here to avoid further crowding of an already demanding detector diagram.
    This structure is inherited from the original $w=8$ 2BGA stabilizer code from which the stairway code is derived.
    }
    \label{fig:detectors}
\end{figure}

In our circuit, there are two types of detectors. Localized detectors, each contained inside one cell, as depicted in Fig.~\ref{fig:detectors}a, and larger detectors, which stem from the original stabilizer code structure, and span several cells, depicted in Fig.~\ref{fig:detectors}b.

The small detector arises directly from the weight-8 spider decomposition in Fig.~\ref{fig:ZX-calculus}e. In this decomposition, one of the measurements is not strictly required; however, including it creates a new detector not present in the original code. This allows for better error detection and, consequently, a higher distance.

The small detector is composed of four measurements, while the big detector is composed of 32 measurements. This number is intimidating and is probably the main cause of the lower threshold compared to several other codes. 
However, the results in Section~\ref{sec:LER_results} reveal that the effect is reasonable and still allow state-of-the-art logical error rate at a low enough physical error rate.

\subsection{Calculating the parameters of a code instance} \label{sec:finding-distance}

In this section, we explain the method we employed to find the number of logical operators and the distance of the code.

While code distance is a well-defined concept for static stabilizer codes—defined as the minimum weight of a logical operator—dynamical codes admit several alternative definitions~\cite{fu_Gottesman2025errorcorrectiondynamicalcodes}. In this work, we consider two distinct notions of distance.

The first is the \emph{embedded distance}, defined as the distance of the embedded homological code. 
In this code, at the timestep subsequent to a pairwise measurement between two qubits, we regard the two measured qubits as one effective qubit, as the measurement projects them into a two-dimensional state.

The second metric is the \emph{circuit-level distance}, $d_\text{circ}$, defined as the minimum number of circuit-level faults required to produce an undetected logical error. Unlike static properties, this distance depends on the choice of noise model. Here, we primarily utilize the EM3 noise model defined in Ref.~\cite{Gidney_2021Honeycomb} and employed in recent studies of hyperbolic Floquet codes~\cite{Higgott_2024hyperbolic, fahimniya2025hyperbolicfloquet}.

Calculating the exact circuit-level distance is computationally intractable for large systems. Therefore, we primarily report an upper bound, obtained by identifying specific sets of native faults that result in a logical error.

\subsubsection{ISG and Non-gauge Logicals}

To find the embedded distance, we try to identify both the instantaneous stabilizer group (ISG) and the non-gauge logicals of the code at every time step. Later, we identify the minimum weight non-gauge logicals using mixed-integer-programming. In this section, we give a detailed description of the first part.




We extract the stabilizer group $\mathcal{S}$ and the logical operators $\mathcal{L}$ by analyzing the evolution of a maximally entangled state rather than arbitrary input states.

Let $A$ denote the register of $n$ physical qubits and $R$ denote a reference register of $n$ ancilla qubits. We initialize the joint system $AR$ in the maximally entangled state $\ket{\Phi} = \bigotimes_{i=1}^n \ket{\phi^+}_{A_i R_i}$, where $\ket{\phi^+}$ is the Bell state $(\ket{00} + \ket{11})/\sqrt{2}$. 
At least one round of the error-correction circuit $U$ is applied solely to register $A$, yielding the state $\ket{\Psi} = (U \otimes I_R) \ket{\Phi}$. This can be efficiently simulated on a clifford simulator.

The state $\ket{\Psi}$ is a stabilizer state described by a symplectic binary matrix $M$ of dimensions $2n \times 4n$ over $\mathbb{F}_2$. The columns of $M$ correspond to the Pauli $X$ and $Z$ components of the system and reference qubits, denoted as $(X_A| Z_A| X_R| Z_R)$. We identify the code properties via Gaussian elimination on $M$:

\begin{enumerate}
    \item \textbf{Stabilizer Generators:} The stabilizers of the code are the operators in the row space of $M$ that act non-trivially on $A$ but as the identity on $R$. We perform Gaussian elimination to row-reduce $M$ such that the columns corresponding to $R$ are cleared. The rows strictly supported on $A$ form the generator set for the code's stabilizer group, $\mathcal{S}$.
    
    \item \textbf{Logical Operators:} The remaining rows, which have non-trivial support on $R$, correspond to the logical operators. Due to the initial entanglement, a Pauli operator on a reference qubit $R_i$ is perfectly correlated with a logical operator on the system $A$. By performing a second Gaussian elimination on the remaining rows (prioritizing the basis of $A$), we isolate the canonical logical operators $\mathcal{L}$ that encode the information protected by the circuit.
\end{enumerate}

This algorithm effectively maps the correlations between the physical and reference systems to the logical structure of the code, providing generators for the ISG and a logical basis for the non-gauge logicals of the code.

To renormalize the ISG and extract the embedded distance, we apply a CNOT gate between qubits measured in the previous round.

\subsubsection{Distance of Dynamical Code}

\begin{figure*}
    \centering
    \def\svgwidth{\linewidth}
    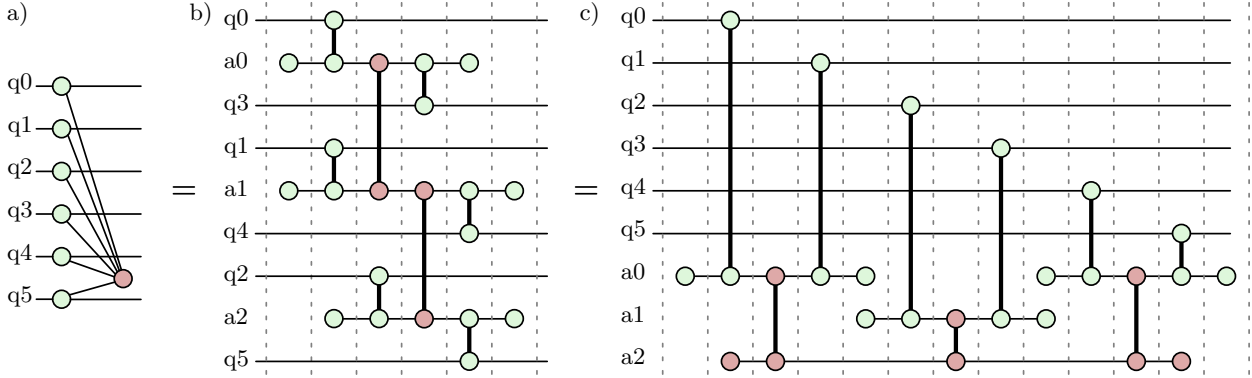
    \caption{Two decompositions of the syndrome extraction circuit of a weight-6 stabilizer into a pairwise measurement schedule. We use both to benchmark our code against bivariate bicycle codes. We note that one ancilla can be dismissed from the second decomposition at the expense of additional idling times.}
    \label{fig:BB-SEC}
\end{figure*}

Integer Linear Programming (ILP) has been widely employed to calculate code distance, as demonstrated in Refs.~\cite{landahl2011ilp_distance_color_code,Bravyi_2024BB_codes}. The standard formulation, detailed in Ref.~\cite{landahl2011ilp_distance_color_code}, searches for the lowest-weight operator that commutes with the stabilizer group but is not itself a stabilizer.

However, dynamical codes (and subsystem codes in general) possess gauge logical operators—operators that commute with the stabilizer group but encode no protected information. These gauge operators often exhibit a lower weight than the logical operators used to store data. Consequently, standard ILP approaches may converge on a gauge operator rather than the relevant logical observable.

To address this, we require a formulation that specifically targets the minimum-weight non-gauge logical operator. As existing methods do not impose this restriction, we solve a modified ILP.

Since we know a set of distinct (non-optimal) logical operators $\{L_q\}$ and a set of (non-optimal) stabilizer generators $\{S_j\}$, we can formulate our minimal-weight logical as
\begin{align}
    L' = \prod  L_q^{\beta_q} \prod  S_j^{\alpha_j}
\end{align}
where $\alpha_q \in \{0,1\}$ and $\beta \in \{0,1\}$ are binary variables to select the stabilizer components and logical components of the minimal-weight logical.

We demand $\sum \beta_q \ge 1$ such that the minimal-weight logical is supported by at least one non-gauge logical.

Marking the support of qubit $i\in \{0,\dots,n\}$ on a stabilizer $j$ as $S_{j,i}$ and on logical $q$ as $L_{q,i}$, and the support on the optimal logical as $L'_i$, we can get the following ILP formulation:
$$\begin{aligned}
\text{min} \quad & \sum_{i=1}^n l_i \\
\text{s.t.} \quad & \sum_{q=0}^{k-1} \beta_q L_{q,i} {+} \sum_{j=1}^m \alpha_j S_{j,i} {-} 2t_i = l_i, ~\forall i \in \{1,\dots,n\} \\
& \sum_{q=0}^{k-1} \beta_q \ge 1 \\
& y_i, \alpha_j, \beta_q \in \{0,1\}, \quad t_i \in \mathbb{Z}_{\ge 0} \;.
\end{aligned}$$
Here, $t_i$ are integer slack variables that allow us to enforce modulo-2 arithmetic within a linear framework over the real numbers.

This ILP can be solved by various ILP solvers~\cite{cbc, gurobi} to get either an optimal answer or an upper bound.

\subsubsection{Circuit-level Distance}

To find the circuit-level distance, we look at the detector-error model~\cite{gidney2021stim} of the circuit.
The circuit-level distance is the minimum weight of an error combination that:
\begin{itemize}
    \item Does not activate any detector.
    \item Flips at least one observable.
\end{itemize}

This can also be formulated in the same manner as above, with stabilizers being detectors, and qubits being error-mechanisms.

\subsection{Finding Code Instances}

To find promising instances of the stairway code family, we perform a numerical search through a generated pool of periodicity vector lists. 

The space of possible periodicity vectors is vast; therefore, we focused on instances with the general structure of the bivariate bicycle code family~\cite{Bravyi_2024BB_codes}. This family has two axis-aligned periodicity vectors, and the rest are non-axial.

Therefore, we generated the search pool out of the template:
\begin{align}
    \left(
    \begin{array}{c c c c c c c c c}
        {[} & x_1 & d_1 & 0 & 0 & 0 & 0 & 0 & {]} \\
        {[} & x_2 & 0 & d_2 & 0 & 0 & 0 & 0 & {]} \\
        {[} & x_3 & a_{1,1} & a_{2,1} & -1 & 0 & 0 & 0 & {]} \\
        {[} & x_4 & a_{2,1} & a_{2,2} & 0 & -1 & 0 & 0 & {]} \\
        {[} & x_5 & a_{3,1} & a_{3,2} & 0 & 0 & -1 & 0 & {]} \\
        {[} & x_6 & a_{4,1} & a_{4,2} & 0 & 0 & 0 & -1 & {]}
    \end{array}
    \right)
\end{align}
with $0 \le a_{i,j} < d_j$, and $x_i$ calculated from $p_i \cdot t = 0$.

Using Eq.~\eqref{eq:number-of-qubits}, we can calculate the number of qubits to be $d_1 d_2 w$.

Scanning over different instances, we identify the notable codes presented in Table \ref{table:codes_intro}. 
The distance upper bounds reported are obtained by running the Gurobi-based ILP decoder for at least 10,000 seconds per instance per distance type, utilizing 32 threads on an AMD EPYC 9R14 processor.


\section{Compiling bivariate bicycles into pairwise measurements} \label{sec:BB-codes}

In addition to hyperbolic Floquet codes, a natural point of comparison for the Stairway codes is the bivariate bicycle code family, provided their syndrome extraction circuits are compiled into pairwise measurements.

While recent works have demonstrated such compilations~\cite{rodatz2024floquetifying_distance_preserving, yichenxu2025spacetimefloquet}, we pursue two different strategies.
The first is analogous to the approach taken in Ref.~\cite{Gidney_2023} for the surface code.
The second is analogous to the approach taken in Ref.~\cite{Chao_2020} for the surface code as well.
To differentiate between the two we refer to them from now on as the ``short" and ``long" constructions, for the left and right construction in the figure, respectively.

We again use the ZX-calculus to break the syndrome extraction circuit into pairwise measurement, as can be seen in Fig.~\ref{fig:BB-SEC}. This requires introducing 3 new ancilla qubits per stabilizer, resulting in $2\times$ increase in the total qubit number relative to the CNOT-based extraction circuit.

Although this compilation is not distance-preserving, we can mitigate the introduction of distance-reducing faults through careful qubit mapping. 

\begin{figure*}
    \centering
    \def\svgwidth{\linewidth}
    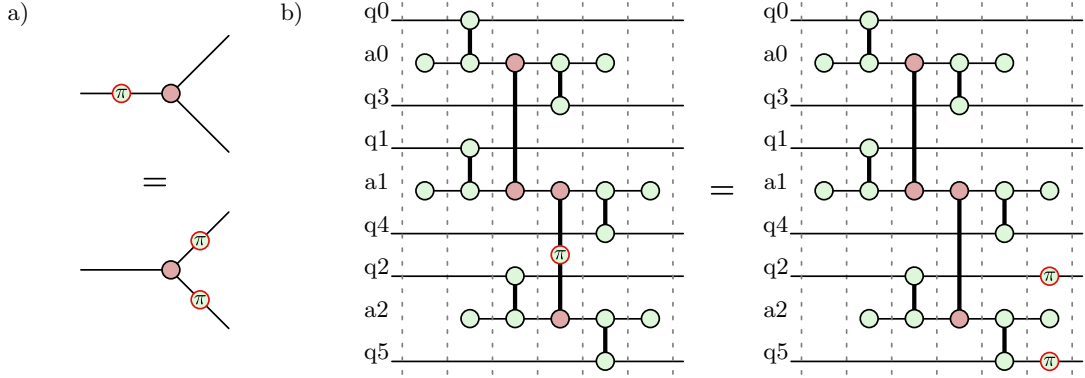
    \caption{Using the ZX-calculus identity presented in (a), we are able to analyze hook errors. (b) An example of a hook error resulting from a measurement error on the $XX$ measurement. The error is equivalent to a Z-spider with the phase of $\pi$, marked by a red border. Using the rule from (a) with the merging and splitting rule from Fig.~\ref{fig:ZX-calculus}e, we can show it is equivalent to two $Z$ errors on data qubits. We present all possible combinations of opposite-type correlated errors in Table~\ref{tab:hook-errors}. }
    \label{fig:BB-hook}
\end{figure*}

Both constructions introduce new ``hook errors''~\cite{Dennis_2002}—faults of the opposite Pauli type that propagate to multiple qubits within a stabilizer. These errors can reduce the circuit-level distance for the opposite basis. 
Hook errors can be analyzed using the ZX-calculus rules, as presented in Fig.~\ref{fig:BB-hook}.
A comprehensive list of hook errors for each construction is provided in Table~\ref{tab:hook-errors}.

\begin{table}[h]
    \centering
    \begin{tabular}{c|c|c}
        
        \textbf{\parbox{2.5cm}{\centering CNOT-based \\ const.}} & 
        \textbf{\parbox{2.5cm}{\centering short \\ const.}} & 
        \textbf{\parbox{2.5cm}{\centering long \\ const.}} \\
        & &
        \\
        \hline
        \hline
        $q_0 ~\&~ q_1 ~\&~ q_2$ & $q_0 ~\&~ q_1 ~\&~ q_3$ & $q_0 ~\&~ q_1 ~\&~ q_2$ \\
        $q_0 ~\&~ q_1$ & $q_0 ~\&~ q_1$ & $q_0 ~\&~ q_1$ \\
        $q_4 ~\&~ q_5$ & $q_4 ~\&~ q_5$ & $q_2 ~\&~ q_3$ \\
        & $q_0 ~\&~ q_3$ & $q_4 ~\&~ q_5$ \\
        & $q_2 ~\&~ q_5$ & $q_2 ~\&~ q_4 ~\&~ q_5$ \\
        & $q_2 ~\&~ q_4$ &  \\
    \end{tabular}
    \caption{Comparison of hook errors combinations between the CNOT and pairwise versions. Each row lists the qubits supporting a correlated error of the opposite type, originating from a single fault during the stabilizer measurement, using the EM3 noise model.}
    \label{tab:hook-errors}
\end{table}

We consider the Gross code~\cite{Bravyi_2024BB_codes} as a test case.
The Gross code is a $[[144,12,12]]$ bivariate bicycle code, which require a total of $288$ qubits to implement with a CNOT-based circuit, and has a circuit-level distance of $\dcirc=10$.
Using the ``short" construction, it can be compiled to use only pairwise measurement and single qubit initialization and measurements, using a total of $576$ qubits with a circuit-level distance of $\dcirc \le8$. 
This might be improved by adding a redundant measurement between a0 and a2, facilitating a new detector. This is left for future work.

For the ``long" construction, we have used the same order given in Ref.~\cite{Bravyi_2024BB_codes}, as if each three pairwise measurements have been two CNOTs, and have observed circuit-level distance of $\dcirc\le 10$, hinting at no degradation in circuit-level distance.

To compare the performance of the Stairway codes, we compare the pairwise-compiled Gross code to the $[[572,14,\le20]]$ Stairway code. We present the logical error rate of both codes in Fig.~\ref{fig:BB-comparison}.

\section{Acknowledgments}

We thank Fernando Brandão for making this work possible. Additionally, we are grateful to Gilad Kishony, as well as Joe Iverson, Przemyslaw Bienias, and others at the AWS Center for Quantum Computing for helpful feedback and discussions.

\section{Data Availability}

The scripts used in this paper, the stim circuits, and the sampled data are available in~\cite{jacoby2026data}.

\appendix

\section{Decoding Time} \label{sec:appndx-decoding-time}

The high-weight detectors of stairway codes have the disadvantage of being computationally expensive for currently popular qLDPC decoders.

\begin{figure}[t]
    \centering
    \includegraphics[width=\linewidth]{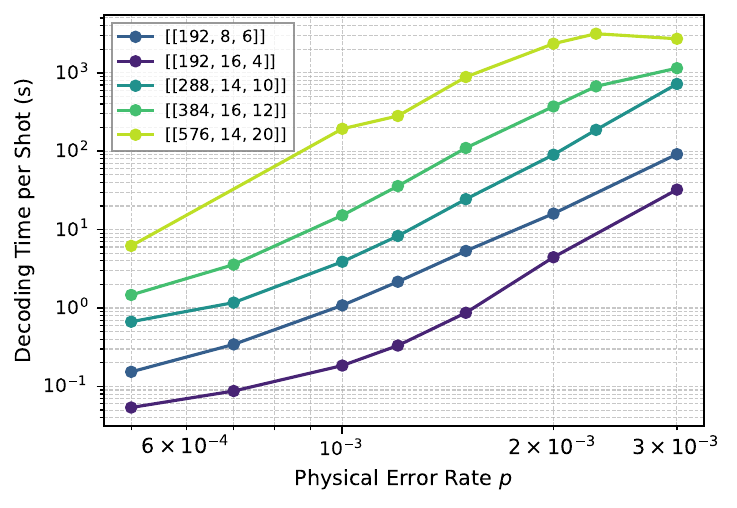}
    \caption{Decoding times for $d$ rounds of memory experiments of the codes listed in Table~\ref{table:codes_intro} using the Tesseract decoder with the ``long-beam" configuration~\cite{beni2025tesseractdecoder}.}
    \label{fig:decoding-time}
\end{figure}

Figure~\ref{fig:decoding-time} presents the decoding time required for $d$ rounds of memory experiment utilizing the Tesseract decoder~\cite{beni2025tesseractdecoder} with the recommended ``long-beam'' configuration. 
We observe that as the code distance increases, the decoding time becomes infeasible, specifically at physical error rates approaching or exceeding the threshold.

Consequently, these results suggest that a tailored decoder is necessary for practical scalability.

To effectively sample the distance-20 code, we have sampled a memory experiment of 6 rounds and renormalized it to 20 rounds. The renormalization method follows the method presented in the Sinter package~\cite{gidney2021stim}. While this is not an exact estimation, we expect it to be adequate.

Given an observed shot error rate $E_{\text{shot}}$ collected over $N_{\text{sample}}$ rounds, we first extract the error rate per-round per-observable, $\varepsilon$, by inverting the cumulative error models:
\begin{equation}
(1 - 2\varepsilon) = \left( 2(1 - E_{\text{shot}})^{1/V} - 1 \right)^{1/N_{\text{sample}}}
\end{equation}
where the term $(1 - E_{\text{shot}})^{1/V}$ isolates the survival probability of a single observable, and the $N_{\text{sample}}$-th root isolates the per-round fidelity assuming logical frame changes are uncorrelated between rounds. The projected error rate for $N_{\text{target}}=20$ rounds is then reconstructed using the forward relation:
\begin{equation}
E_{\text{target}} = 1 - \left( \frac{1 + (1 - 2\varepsilon)^{N_{\text{target}}}}{2} \right)^V
\end{equation}

\section{Logical Error Rate Fit}

To estimate the logical error rates at $p=10^{-3}$ and the pseudo-threshold value, we have fitted the Monte-Carlo samplings using the ansatz from Ref.~\cite{Bravyi_2024BB_codes}
\begin{align} \label{eq:fit-eq}
    p_L=p^{\dcirc'}/2e^{c_0+c_1 p + c_2 p^2},
\end{align}
with $\dcirc'$ being the upper bound on $\dcirc$ from Table~\ref{table:codes_intro} and $c_0, c_1,c_2$ are fitting parameters.

To mitigate artifacts arising from decoder queue exhaustion at high physical error rates, we included only data points below the threshold.

\begin{figure}
    \centering
    \includegraphics[width=\linewidth]{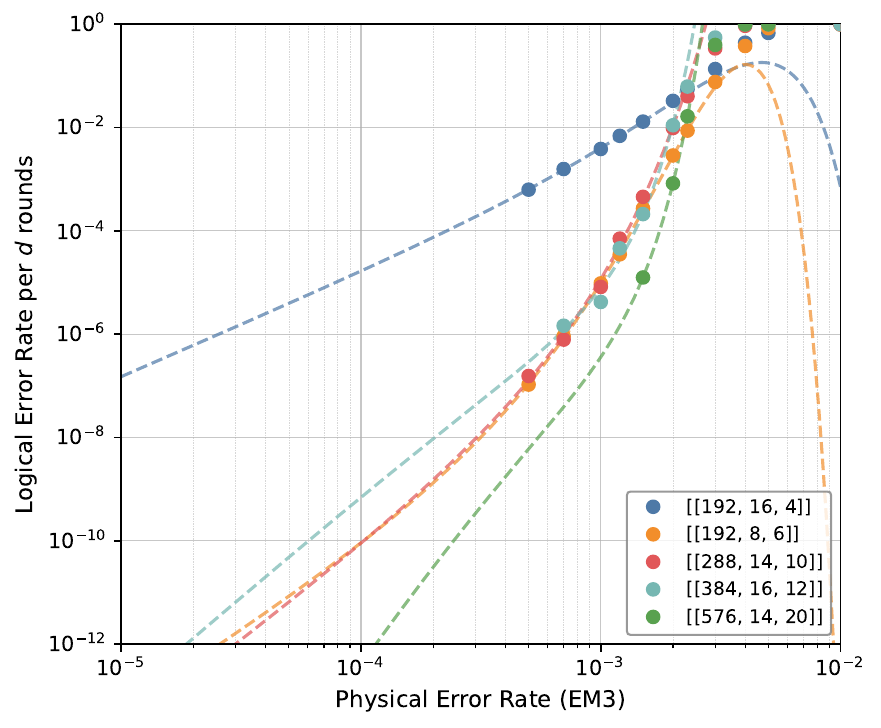}
    \caption{The sampled data fitted to Eq. \eqref{eq:fit-eq}. Only points below the pseudo-threshold are used for the fit.}
    \label{fig:error-rate-fit}
\end{figure}

The fitting results are presented in Fig.~\ref{fig:error-rate-fit}.

\section{Embedded distance and pairwise measurements using CNOT gates} \label{sec:embedded-distance-for-CNOT}

\begin{figure}[b]
    \centering
    \def\svgwidth{0.7\linewidth}
\begingroup%
  \makeatletter%
  \providecommand\color[2][]{%
    \errmessage{(Inkscape) Color is used for the text in Inkscape, but the package 'color.sty' is not loaded}%
    \renewcommand\color[2][]{}%
  }%
  \providecommand\transparent[1]{%
    \errmessage{(Inkscape) Transparency is used (non-zero) for the text in Inkscape, but the package 'transparent.sty' is not loaded}%
    \renewcommand\transparent[1]{}%
  }%
  \providecommand\rotatebox[2]{#2}%
  \newcommand*\fsize{\dimexpr\f@size pt\relax}%
  \newcommand*\lineheight[1]{\fontsize{\fsize}{#1\fsize}\selectfont}%
  \ifx\svgwidth\undefined%
    \setlength{\unitlength}{167.39512406bp}%
    \ifx\svgscale\undefined%
      \relax%
    \else%
      \setlength{\unitlength}{\unitlength * \real{\svgscale}}%
    \fi%
  \else%
    \setlength{\unitlength}{\svgwidth}%
  \fi%
  \global\let\svgwidth\undefined%
  \global\let\svgscale\undefined%
  \makeatother%
  \begin{picture}(1,0.52529052)%
    \lineheight{1}%
    \setlength\tabcolsep{0pt}%
    \put(0.49736308,0.45173706){\color[rgb]{0,0,0}\makebox(0,0)[t]{\lineheight{1.25}\smash{\begin{tabular}[t]{c}$ZZ$ measurement\end{tabular}}}}%
    \put(0,0){\includegraphics[width=\unitlength,page=1]{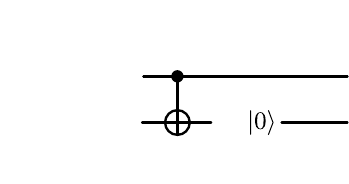}}%
    \put(0.59127248,0.08127139){\color[rgb]{0,0,0}\makebox(0,0)[lt]{\lineheight{1.25}\smash{\begin{tabular}[t]{l}$b$\end{tabular}}}}%
    \put(0,0){\includegraphics[width=\unitlength,page=2]{CNOT_pairwise_meas.pdf}}%
    \put(0.13674257,0.01010892){\color[rgb]{0,0,0}\makebox(0,0)[t]{\lineheight{1.25}\smash{\begin{tabular}[t]{c}$b$\end{tabular}}}}%
    \put(0,0){\includegraphics[width=\unitlength,page=3]{CNOT_pairwise_meas.pdf}}%
  \end{picture}%
\endgroup%

    \caption{Pairwise $ZZ$ measurement can be implemented using two CNOTs, a measurement and an initialization. This implementation is non-fault tolerant, as errors can spread to both qubits. However, when considering the \emph{embedded distance} of a Floquet code, it is equivalent as the ISG distance when implementing the measurements with CNOTs.}
    \label{fig:CNOT-meas}
\end{figure}

Pairwise measurements can be implemented without an ancilla qubit by utilizing two CNOT gates alongside mid-circuit measurement and initialization, as illustrated in Fig.~\ref{fig:CNOT-meas}.

Consequently, a Floquet code relying on a pairwise measurement schedule can be directly adapted for CNOT-based architectures. While this approach incurs a circuit-level performance penalty—since each pairwise measurement expands into four distinct operations—it yields valuable insights into the code's embedded distance, presented in Section~\ref{sec:distance-results}.

Because single-qubit errors during a CNOT-based implementation can propagate into correlated Pauli errors on the two measured qubits, the embedded distance of the Floquet code is equivalent to the Instantaneous Stabilizer Group (ISG) distance of the code on this architecture (i.e., the code distance under a phenomenological noise model).

Under this model, The $[[288,14,9\le d\le10]]$ Stairway code directly compares to the Gross code including ancillas, which has the parameters $[[288,12,\le10]]$\footnote{The distance reported for both of those codes includes circuit-level features like hook errors, but does not include correlated two-qubit errors after CNOTs.}.

\bibliographystyle{apsrev4-2}
\bibliography{bib}

\end{document}